\documentclass[useAMS,usenatbib]{mn2e}
\usepackage[a4paper,margin=2cm]{geometry}
\usepackage[pdftex]{graphicx}
\pdfoutput=1
\usepackage{color}
\usepackage{aas_macros}
\usepackage{lastpage}
\usepackage{comment}
\usepackage{amsmath}
\usepackage{type1cm}
\usepackage{eso-pic}

\def\lesssim{\mathrel{\hbox{\rlap{\hbox{\lower4pt\hbox{$\sim$}}}\hbox{$<$}}}}
\def\gtrsim{\mathrel{\hbox{\rlap{\hbox{\lower4pt\hbox{$\sim$}}}\hbox{$>$}}}}
\newcommand{\ltaraw}{$\; \buildrel < \over \sim \;$}
\newcommand{\lta}{\lower.5ex\hbox{\ltaraw}}
\newcommand{\gtaraw}{$\; \buildrel > \over \sim \;$}
\newcommand{\gta}{\lower.5ex\hbox{\gtaraw}}
\def\lsim{\mathrel{\rlap{\lower4pt\hbox{\hskip1pt$\sim$}}
    \raise1pt\hbox{$<$}}}                

\newcommand{\ie}{{\it i.e.}}
\newcommand{\eg}{{\it e.g.}}

\newcommand{\etal}{{\it et~al.}}

\newcommand{\LCDM}{$\Lambda$CDM}

\def\apjl{ApJ}%
\def\apjs{ApJS}%
\def\ao{Appl.~Opt.}%
\def\aap{A\&A}%
\newcommand{\zeff}{{$z_{\rm eff}$}}

\title[WiggleZ: Probing the Epoch of Radiation Domination]{The WiggleZ Dark Energy Survey: Probing the Epoch of Radiation Domination using Large Scale Structure}

\author[Poole et al.]{\parbox[t]{\textwidth}{
    Gregory B.\ Poole$^{1,2}$\footnotemark,
    Chris Blake$^1$,
    David Parkinson$^3$, \\
    Sarah Brough$^4$, 
    Matthew Colless$^4$, 
    Carlos Contreras$^1$,
    Warrick Couch$^1$, \\ 
    Darren J. Croton$^1$, 
    Scott Croom$^5$, 
    Tamara Davis$^3$, 
    Michael J.\ Drinkwater$^3$, \\
    Karl Forster$^6$, 
    David Gilbank$^7$,  
    Mike Gladders$^8$, 
    Karl Glazebrook$^1$, 
    Ben Jelliffe$^5$, 
    Russell J.\ Jurek$^9$, 
    I-hui Li$^{10}$, 
    Barry Madore$^{11}$, 
    D.\ Christopher Martin$^6$, \\
    Kevin Pimbblet$^{12}$, 
    Michael Pracy$^{1,13}$, 
    Rob Sharp$^{4,13}$,
    Emily Wisnioski$^{1,14}$, \\
    David Woods$^{15}$, 
    Ted K.\ Wyder$^6$ and H.K.C. Yee$^{10}$} \\ \\
  $^1$ Centre for Astrophysics \& Supercomputing, Swinburne University of Technology, P.O. Box 218, Hawthorn, VIC 3122, Australia \\ 
  $^2$ School of Physics, University of Melbourne, Parksville, VIC 3010, Australia \\
  $^3$ School of Mathematics and Physics, University of Queensland, Brisbane, QLD 4072, Australia \\
  $^4$ Australian Astronomical Observatory, P.O. Box 915, North Ryde, NSW 1670, Australia \\ 
  $^5$ Sydney Institute for Astronomy, School of Physics, University of Sydney, NSW 2006, Australia \\ 
  $^6$ California Institute of Technology, MC 278-17, 1200 East California Boulevard, Pasadena, CA 91125, United States \\ 
  $^7$ South African Astronomical Observatory, PO Box 9 Observatory, 7935 South Africa \\
  $^8$ Department of Astronomy and Astrophysics, University of Chicago, 5640 South Ellis Avenue, Chicago, IL 60637, United States \\ 
  $^{9}$ Australia Telescope National Facility, CSIRO, Epping, NSW 1710, Australia \\ 
  $^{10}$ Department of Astronomy and Astrophysics, University of Toronto, 50 St.\ George Street, Toronto, ON M5S 3H4, Canada \\
  $^{11}$ Observatories of the Carnegie Institute of Washington, 813 Santa Barbara St., Pasadena, CA 91101, United States \\ 
  $^{12}$ School of Physics, Monash University, Clayton, VIC 3800, Australia \\ 
  $^{13}$ Research School of Astronomy \& Astrophysics, Australian National University, Weston Creek, ACT 2600, Australia \\ 
  $^{14}$ Max Planck Institut f\"{u}r extraterrestrische Physik, Giessenbachstra$\beta$e, D-85748 Garching, Germany\\
  $^{15}$ Department of Physics \& Astronomy, University of British Columbia, 6224 Agricultural Road, Vancouver, BC V6T 1Z1, Canada} 
\date{draft version \today}
\pagerange{\pageref{firstpage}--\pageref{lastpage}} \pubyear{2012}

\begin{document}

\label{firstpage}

\maketitle

\begin{abstract}
We place the most robust constraint to date on the scale of the turnover in the cosmological matter power spectrum using data from the WiggleZ Dark Energy Survey.  We find this feature to lie at a scale of $k_0{=}0.0160^{+0.0041}_{-0.0035}$ [$h$/Mpc] ($68$\% confidence) for an effective redshift of \zeff${=}0.62$ and obtain from this the first-ever turnover-derived distance and cosmology constraints: a measure of the cosmic distance-redshift relation in units of the horizon scale at the redshift of radiation-matter equality ($r_{\rm H}$) of $D_{\rm V}(z_{\rm eff}{=}0.62){/}r_{\rm H}{=}18.3^{+6.3}_{-3.3}$ and, assuming a prior on the number of extra relativistic degrees of freedom $N_{\rm eff}{=}3$, constraints on the cosmological matter density parameter $\Omega_{\rm M}h^2{=}0.136^{+0.026}_{-0.052}$ and on the redshift of matter-radiation equality $z_{\rm eq}{=}3274^{+631}_{-1260}$.  We stress that these results are obtained within the theoretical framework of Gaussian primordial fluctuations and linear large-scale bias.  With this caveat, all results are in excellent agreement with the predictions of standard \LCDM\ models.  Our constraints on the logarithmic slope of the power spectrum on scales larger than the turnover is bounded in the lower limit with values only as low as $-1$ allowed, with the prediction of $P(k){\propto}k$ from standard \LCDM\ models easily accommodated by our results.  Lastly, we generate forecasts to estimate the achievable precision of future surveys at constraining $k_{\rm 0}$, $\Omega_{\rm M} h^2$, $z_{\rm eq}$ and $N_{\rm eff}$.  We find that the Baryon Oscillation Spectroscopic Survey (BOSS) should substantially improve upon the WiggleZ turnover constraint, reaching a precision on $k_{\rm 0}$ of ${\pm}9$\% ($68$\% confidence),  translating to precisions on $\Omega_{\rm M} h^2$ and $z_{\rm eq}$ of ${\pm}10$\% (assuming a prior $N_{\rm eff}{=}3$) and on $N_{\rm eff}$ of ${}^{+78}_{-56}$\% (assuming a prior $\Omega_{\rm M} h^2{=}0.135$).  This represents sufficient precision to sharpen the constraints on $N_{\rm eff}$ from WMAP, particularly in its upper limit.  For Euclid, we find corresponding attainable precisions on $(k_{\rm 0}, \Omega_{\rm M} h^2, N_{\rm eff})$ of $(3,4,{}^{+17}_{-21})$\%.  This represents a precision approaching our forecasts for the Planck Surveyor.

\end{abstract}


\section{Introduction}\label{sec-intro} 
\renewcommand{\thefootnote}{\fnsymbol{footnote}}
\setcounter{footnote}{1}
\footnotetext{E-mail: gpoole@unimelb.edu.au}

Recent decades have witnessed an incredible refinement of our cosmological model with significant advancements in the volume and redshift coverage of galaxy surveys, and in the methods of their analysis, playing a pivotal role.  Such advances have opened up entirely new avenues of investigation, notably the use of features in the distribution of galaxies as standard rulers for direct geometric mapping of the Universe's expansion history.

The most investigated of these features are those induced by ``Baryon Acoustic Oscillations'' (BAOs) in the early-Universe's photon-baryon fluid.  The precise measurement across cosmic time of the BAO scale in the distribution of galaxies -- a weak harmonic ripple with a comoving fundamental scale of ${\sim}105$ [Mpc/$h$] -- has been a major driver of recent and future galaxy redshift surveys and has received a great deal of attention in recent literature \citep[e.g.][]{Cole:2005p1578,Eisenstein:2005p1275,Blake:2011a,Beutler:2011p1579,Blake:2011p1580,Anderson:2012p1581}.  

Far less studied, the ``turnover'' in the galaxy power spectrum also represents a standard-ruler in the observed distribution of galaxies.  This feature, which is predicted to manifest on comoving scales of ${\sim}400$ [Mpc/$h$] (or corresponding wavenumber $k_{\rm 0}{\sim}0.016$ [$h$/Mpc]), was established at the epoch when the dynamics of the Universe transitioned from being dominated by relativistic material (photons and neutrinos) to being dominated by matter (dark and baryonic).  This transition occurred because  the mass-energy density of relativistic and non-relativistic materials decline differently with $a$, the expansion factor of the Universe.  The particle density of both decline as $\rho \propto a^{-3}$ but the per-particle energy of relativistic material also decreases with $a$ due to redshift effects.

Previous to this transition from radiation domination to matter domination (at redshift $z_{\rm eq}$), oscillations smaller than the horizon in the strongly coupled matter-radiation field were suppressed by the effects of radiation pressure while causally disconnected fluctuations larger than the horizon collapsed unhindered \citep[see][]{Eisenstein:1998p1770}.  As a result, the scale-free primordial matter power spectrum believed to have emerged from inflation, $P(k){\propto}k^{n_{\rm s}}$ with $n_{\rm s}{\sim}1$, became distorted such that $P(k)$ became a  \emph{decreasing} function of $k$ at small scales with a limiting behaviour $P(k){\propto}k^{-3}$.  At the turnover scale, a peak in $P(k)$ arose separating scales where $P(k)$ increased with $k$ from those which decreased with $k$.  

While the size of the horizon grew during the epoch of radiation domination, the scale of the turnover shifted ever larger to smaller values of $k_{\rm 0}$ until the epoch of matter domination commenced at redshift $z_{\rm eq}$ and the suppression of small scale fluctuations ceased.  From this point forward, all scales grew independently and by the same fractional amount while in the linear regime, and the comoving scale of the turnover became fixed.  The exact size of this scale and the time of matter-radiation equality was influenced by the matter and radiation mass-energy densities; $\Omega_{\rm M} h^2$ and $\Omega_{\rm r} h^2$, the latter being set by the effective number of extra relativistic degrees-of-freendom ($N_{\rm eff}$) and by the photon mass-energy density ($\Omega_{\gamma} h^2$) which is well determined from observations of the temperature of the Cosmic Microwave Background (CMB).  

Analyses of the CMB have placed the most powerful constraints to date on parameters describing the epoch of matter-radiation equality by measuring $z_{\rm eq}$, one of the ``fundamental observables'' of the CMB \citep{Komatsu:2009}.  This measurement is made possible because of early integrated Sachs-Wolfe effects which manifest in the ratio of power between the 1$^{\rm st}$ and 3$^{\rm rd}$ peaks in the angular power spectrum.  The latest WMAP results find $z_{\rm eq}=3145^{+140}_{-139}$ \citep{Komatsu:2011p1737}.

Studies of the matter power spectrum on scales of the turnover can provide an additional window into physical processes operating at (or even prior to) the epoch of matter-radiation equality; a time before even the epoch of recombination observed by CMB experiments.   While the precision of the turnover scale as a standard ruler is reduced by the fact that it less sharply defined (fractionally speaking) than the BAO scale, its study benefits from lying firmly in the linear regime at all redshifts, easing complications which arise from non-linear structure formation and redshift-space distortions.  As a result, structure on scales of the turnover are sensitive to non-Gaussian processes during inflation permitting interesting new studies of early-Universe physics \citep{Feldman:1994p1575,Durrer:2003p1569}.  For instance, scale-dependent bias effects are expected for galaxy samples with biases deviating from unity \citep[\eg][]{Dalal:2008p1750}, providing a rare opportunity to study inflation.  Even in the case of Gaussian fluctuations, informative scale-dependent effects may be present on scales beyond the turnover \citep{Yoo:2010p1773}.

The sensitivity of turnover scales to $N_{\rm eff}$ is also of interest given the conflict of several recent CMB studies with conventional theoretical expectations.  For the standard \LCDM\ model, the expected value is $N_{\rm eff}{\sim}3.046$ \citep[][\ie\ slightly larger than $3$ owing to the fact that neutrino decoupling was not instantaneous]{Mangano:2005p1731}.  However, recent high spatial resolution CMB measurements from the ACT and SPT observatories \citep{Dunkley:2011p1732,Keisler:2011p1733} suggest a significantly higher result of $N_{\rm eff}{\sim}4$ \citep{Hou:2011p1735,Smith:2012p1736,Calabrese:2012p1730} leading several researchers to explore a class of ``dark radiation'' models \citep[\eg][]{Archidiacono:2011p1734}.

The primary challenge to studies of the turnover, and the reason for its scant study to date, is the large volumes one must uniformly probe to detect it with any precision.  While the fluctuations involved may lie on ${\sim}400$ [Mpc/$h$] scales, many modes on these scales must be enclosed by a survey's volume to permit a statistically significant analysis.  Furthermore, subtle errors in the calibration of a survey's selection function across its volume can easily lead to significant systematic power spectrum distortions on a survey's largest scales.  This can be particularly serious for angular (\ie\ 2D rather than 3D) measurements in imaging surveys which are much more sensitive to angular systematics in photometry and/or target selection \citep[\eg][]{Ross:2011p1778,Ross:2012p1753}.

Despite these challenges, several attempts have been made to measure the turnover or to study the matter power-spectrum on its scales.  The earliest attempts were performed using the APM galaxy survey \citep{Baugh:1993p1091,Baugh:1994p1092} or through studies of the distribution of optically-selected galaxy clusters \citep{Peacock:1992p1339,Scaramella:1993p1573,Einasto:1993p1094,Tadros:1998p1572,Einasto:1999p1097}.  These studies have tended to observe turnovers at scales of ${\sim}100-200$ [Mpc/$h$], which is quite discrepant with a wide range of modern cosmological probes and analyses.  In all cases, systematic effects induced by survey selection were found to dominate on turnover scales or rigorous studies of such issues were not performed.  More recently, observations of quasars \citep{Outram:2003p1765} and luminous red galaxies (LRGs) in the Sloan Digital Sky Survey \citep[SDSS;][]{York:2000p1772,Padmanabhan:2007p1128} have managed to measure power on scales reaching to those of the turnover but relatively little analysis using this information has been performed.

In this paper we present the most robust measurement of the turnover scale to date using data from the WiggleZ Dark Energy Survey \citep{Drinkwater:2010}.  WiggleZ is a large-scale galaxy redshift survey conducted with the AAOmega multi-object spectrograph on the Anglo-Australian Telescope at Siding Spring Observatory.  WiggleZ was designed to study the effect of dark energy on the Universe's expansion history and on the growth of cosmological structures across an unprecedented period of cosmic history.  The primary science drivers for the survey have been the measurement of the BAO scale \citep[\eg][]{Blake:2011a}, of the growth of cosmological structure \citep[\eg][]{Blake:2011b} and measurement of the neutrino mass \citep[\eg][]{RiemerSrensen:2012p1754} using the clustering pattern of UV-selected galaxies, as well as studies of the Universe's most actively star forming galaxies out to a redshift of 1.2 \citep[\eg][]{Wisnioski:2011,Li:2012p1729}.  

In this work we treat the dominant systematic uncertainty of our study coming from convolution effects of the survey selection function and the subsequent extraction of the turnover scale in a statistically rigorous way.  We also perform forecasts for the ongoing and future surveys of the Baryon Oscillation Spectroscopic Survey \citep[BOSS; ][]{Eisenstein:2011p1756} and Euclid \citep{Laureijs:2011p1755}.

Our analysis is presented purely within the theoretical framework of Gaussian primordial fluctuations.  Models which permit non-Gaussian primordial fluctuations provide for a wide range of large-scale clustering behaviours and we will leave their exploration for future work.

In Section \ref{sec-observations} we present our method of constructing a galaxy power spectrum from the WiggleZ dataset which is optimised for studies of the turnover.  In Section \ref{sec-analysis} we extract the turnover scale from the WiggleZ observations and present the resulting distance measurement to $z{=}0.62$ and constraints on $\Omega_{\rm M} h^2$ and $z_{\rm eq}$.  In Section \ref{sec-forecasts} we perform forecasts for future surveys (highlighting results for BOSS and Euclid), estimating the volume dependance of turnover scale measurement precision and of resulting cosmology constraints.  We also compute the constraints in the $N_{\rm eff}{-}\Omega_M h^2$ plane from the CMB observations of WMAP and forecast the constraints achievable by the Planck Surveyor.  Lastly, we present some discussion and our conclusions in Section \ref{sec-summary}.

Our choice of fiducial cosmology throughout will be a standard \LCDM\ model with $\Omega_{\rm M}{=}0.27$, $\Omega_\Lambda{=}0.73$ and $h{=}0.7$, unless otherwise stated.

\section{Observations}\label{sec-observations}

\begin{table*}
\caption[WiggleZ power spectrum parameters]{A summary of parameters relevant to our calculation of the WiggleZ galaxy power spectrum for each observed region.  The values of ($L_{\rm x}$,$L_{\rm y}$,$L_{\rm z}$) specify the dimensions of the cuboid used to enclose each region and ($n_{\rm x}$,$n_{\rm y}$,$n_{\rm z}$) specify the corresponding grid dimensions of the cuboid.  Volume and $N_{\rm gal}$ are the volume and number of galaxies contributing to our measurement in each region (spanning the redshift range $[z_{\rm min},z_{\rm max}]=[0.4,0.8]$) respectively while $n$ is the resulting average number density of galaxies in each region.\label{table-Pk_parameters}}
\begin{tabular}{rccccccccccc}
\hline
Region & $L_{\rm x}$ [${\rm Mpc}/h$] & $L_{\rm y}$ [${\rm Mpc}/h$] & $L_{\rm z}$ [${\rm Mpc}/h$] & $n_{\rm x}$ & $n_{\rm y}$ & $n_{\rm z}$ & Volume [$({\rm Gpc}/h)^3$] & $N_{\rm gal}$ & $n$ [$({\rm Mpc}/h)^{-3}$] \\
\hline
 $9$-hour  & $899.4$ & $520.7$ & $315.5$ & $256$ & $128$  & $64$ & $0.148$   & $18978$ & $1.28\times10^{-4}$ \\
$11$-hour & $899.5$ & $520.7$ & $318.9$ & $256$ & $128$  & $64$ & $0.149$   & $20170$ & $1.35\times10^{-4}$ \\
$15$-hour & $907.7$ & $694.7$ & $353.4$ & $256$ & $128$  & $64$ & $0.223$   & $30015$ & $1.35\times10^{-4}$ \\
$22$-hour & $894.4$ & $335.1$ & $338.6$ & $256$ & $64$    & $64$ & $0.101$   & $16146$ & $1.59\times10^{-4}$ \\
 $1$-hour  & $891.8$ & $300.7$ & $252.4$ & $256$ & $64$    & $64$ & $0.0677$ & $8304$   & $1.23\times10^{-4}$ \\
 $3$-hour  & $893.1$ & $313.3$ & $305.6$ & $256$ & $64$    & $64$ & $0.0855$ & $10241$ & $1.20\times10^{-4}$ \\
 $0$-hour  & $891.6$ & $241.1$ & $297.8$ & $256$ & $64$    & $64$ & $0.0640$ & $7409$   & $1.16\times10^{-4}$ \\
\hline
\end{tabular}
\end{table*}

In this section we present our construction of the WiggleZ power spectrum for use in our turnover analysis.  This will include our methods of deconvolving the effects of the survey selection function, of coadding the power spectra observed in seven independent WiggleZ survey regions and our treatment of the survey's radial selection function; the dominant source of systematic uncertainty in this current analysis. 

\subsection{The WiggleZ Power Spectrum}

To construct a power spectrum for this study, we followed the general approach described by \citet[][see Section 3.1]{Blake:2010p74}.  To summarise, we used the optimal weighting scheme of \citet[][FKP]{Feldman:1994p1575}, converting redshifts to distances using our fiducial \LCDM\ cosmological model. We individually enclosed the survey cones of each region within cuboids of sides ($L_{\rm x}$,$L_{\rm y}$,$L_{\rm z}$) and map each region's observed galaxy distribution onto a grid with dimensions ($n_{\rm x}$,$n_{\rm y}$,$n_{\rm z}$) using the ``nearest grid point'' assignment scheme.  Denoting the resulting distribution as $n(\vec{x})$, we then applied a Fast Fourier Transform to the resulting grid, weighted in accordance with the method described in FKP (as given by Eqn. 9 of Blake et al. 2010, using a weighting factor more appropriate for our turnover analysis of $P_0=20000$ $\left[({\rm Mpc}/h)^3\right]$), to produce  $n(\vec{k})$.  The corresponding values for each WiggleZ region are presented in Table \ref{table-Pk_parameters}.  In later sections we will require the covariance between the bins in our WiggleZ $P(k)$.  This was also determined following the procedure of FKP as presented in Section 3.1 (Eqn. 20) of \citet{Blake:2010p74}.  We note that increasing the characteristic amplitude used in the FKP weighting from $P_0=2500$ $\left[({\rm Mpc}/h)^3\right]$ used previously for WiggleZ BAO studies could amplify systematics originating from low-density regions, but in practice we find the consequences are not significant for this dataset.

Both the FKP weighting and the conversion of $n(\vec{k})$ to our final estimate of the power spectrum, $P(k)$, depend critically on the survey selection function, $W(\vec{x})$.  This function expresses the expected mean density of galaxies with spectroscopic redshifts at position $\vec{x}$, given the angular and luminosity survey selection criteria.  Again, we follow the procedure of \citet[][see Section 2]{Blake:2010p74} to determine the angular and radial contributions to $W(\vec{x})$.  While angular effects are estimated in precisely the same manner as described in \citet{Blake:2010p74} -- accounting for effects such as the coverage mask, spatial variations in dust extinction and radial completeness of each individual 2df pointing contributing to the survey -- our estimation of radial contributions to the survey selection function required some modification which is described below in Section \ref{sec-radial_selection}.

While aliasing from the assignment scheme only affects the power spectrum on scales far smaller than the turnover, we nevertheless correct for this effect using the method described by \citet{Jing:2005p1576}.  Furthermore, since the survey target density is obtained from the survey itself, we correct for a potential large-scale $P(k)$ bias in a manner analogous to the integral constraint of correlation functions \citep[see Eqn. 25 of][for example]{Peacock:1991p1779} by applying a boost to the measured $P(k)$ near $k{=}0$, generated using the Fourier transform of the window function.

Lastly, to ensure a power spectrum optimised for studying scales around the turnover, we have made the following three adjustments to the procedure described in \citet{Blake:2010p74}.

\subsubsection{Redshift Range}

The redshift range used in our analysis has two competing effects on our analysis.  By increasing the range used we increase the effective volume of our measurement, increasing the number of modes probed by our study at scales of the turnover and increasing the signal of our volume limited measurement.  Unfortunately, doing so also leads to a less plane-parallel survey geometry, increasing the covariance induced by the survey selection function on our largest-scale modes.

We have performed our analysis on several redshift ranges $[z_{\rm min},z_{\rm max}]$, varying $z_{\rm min}$ from 0.3 to 0.4 and $z_{\rm max}$ from 0.8 to 0.9.  While the choice over these ranges does not have a strong effect on our results, we have found the range $[0.4,0.8]$ to yield slightly optimal results since it maintains minimal covariance between neighbouring $P(k)$ bins while providing a near-maximally precise measurement.  Hence, we will use this range throughout in our analysis. 
 
\subsubsection{Choice of Fourier Binning}

We perform all our analysis on power spectra binned by wave-number ($k$).  Our choice of binning also introduces two competing effects.  By increasing the size of our power spectra bins, we increase the signal and reduce the covariance of neighbouring bins, increasing the precision of our power spectra.  Unfortunately, we are trying to fit to a feature at small values of $k$ and the use of bins that are too large will prevent us from resolving it.  We have experimented with a variety of bin sizes ranging from $\delta k=0.005$ [$h$/Mpc] to $\delta k=0.02$ [$h$/Mpc] and have found the use of $\delta k=0.005$ [$h$/Mpc] to be optimal.  We deemed further reduction of this binning unwarranted given the low signal at this point for modes larger than the turnover in several regions.  In all cases, the medians of the $|\vec{k}|$ values contributing to each bin are used to represent their positions.

\subsubsection{Radial Selection Function}\label{sec-radial_selection}

At the mean of our chosen redshift range, the maximum scale probed by WiggleZ in directions transverse to the line of sight is $\sim 500$ [Mpc/$h$] while the comoving distance along the line of sight from $z_{\rm min}{=}0.4$ to $z_{\rm max}{=}0.8$ is ${\sim}900$ [Mpc/$h$].  As a result, most of the information in WiggleZ on scales of the turnover is contained within radial modes.  Consequently, our estimate of the WiggleZ radial selection function is a significant source of systematic uncertainty in our present analysis.

We have examined several approaches to determining the WiggleZ radial selection function based on the approach presented in \citet[][see Section 2.5]{Blake:2010p74}.  Summarising briefly: we fit a smooth analytic function to the observed redshift distribution $N(z)$ for each observing priority band (WiggleZ observations were prioritised by magnitude, with apparently-faint galaxies given highest priority and each band representing an equal interval of the range $20.0 \le r \le 22.5$).  In this procedure we must choose a functional form with which to parameterise the observed WiggleZ $N(z)$.  In past efforts, this procedure has been performed on each of the seven WiggleZ survey regions \emph{independently} using a polynomial of order dynamically chosen to be that above which the reduced-$\chi^2$ statistic does not decrease.

We have reexamined this procedure using several approaches and present three here: the standard polynomial fit applied to each region independently (\ie\ the standard approach of past work; denoted the ``polynomial $N(z)$'' method), a cubic spline similarly fit to each region separately (denoted the ``spline $N(z)$'' method) and a polynomial fit to the total sum of $N(z)$ for all Northern Galactic Pole (NGP) fields jointly and all Southern Galactic Pole (SGP) fields jointly (denoted the ``joint polynomial $N(z)$'' method; note, due to differing optical selection, the WiggleZ NGP and SGP fields have differing radial selection functions).  The motivation of this last method is to reduce the effects of cosmic variance by combining regions but carries the risk of being susceptible to region-to-region systematics.  The influence of this choice on our resulting WiggleZ $P(k)$ is shown in Figure \ref{fig-Pk_deconvolved} where we show the deconvolved WiggleZ power spectrum for the whole survey sample (see Section \ref{sec-Pk_deconvolved} below for details).  As noted in \citet{Blake:2010p74}, only results for $k<0.03$ [$h$/Mpc] are significantly affected by the details of our radial selection estimation, leaving the scales of the BAO unaffected.

We have carried our analysis through with each of these methods and find relatively little effect on our turnover constraints and no qualitative change to the conclusions of this study.  The challenge in the procedure of determining $N(z)$ for this purpose is to choose a functional form nuanced enough to capture the selection induced by the survey strategy and telescope operations but smooth enough not to fit to (and hence, remove) the real structure we seek to measure.  Hence, for this reason we favour the joint polynomial $N(z)$ and for all subsequent analysis we quote results derived using this approach.

Lastly, our method includes a full $k$-dependent correction for misidentified redshifts (referred to as redshift ``blunders'') using the survey simulations described in section 3.2 of \citet{Blake:2010p74}.  This includes blunders both from constant-multiple shifts due to emission-line mis-identifications and a continuous range of mis-identified sky lines into a lognormal simulation including the full selection function.  The redshift blunder model is based on thousands of repeat redshifts obtained during the survey.  We find this correction to be small compared to the P(k) uncertainty for the scales relevant to our turnover analysis.

\subsection{Deconvolution of the Survey Selection Function}\label{sec-Pk_deconvolved}

\begin{figure}
\includegraphics[width=90mm]{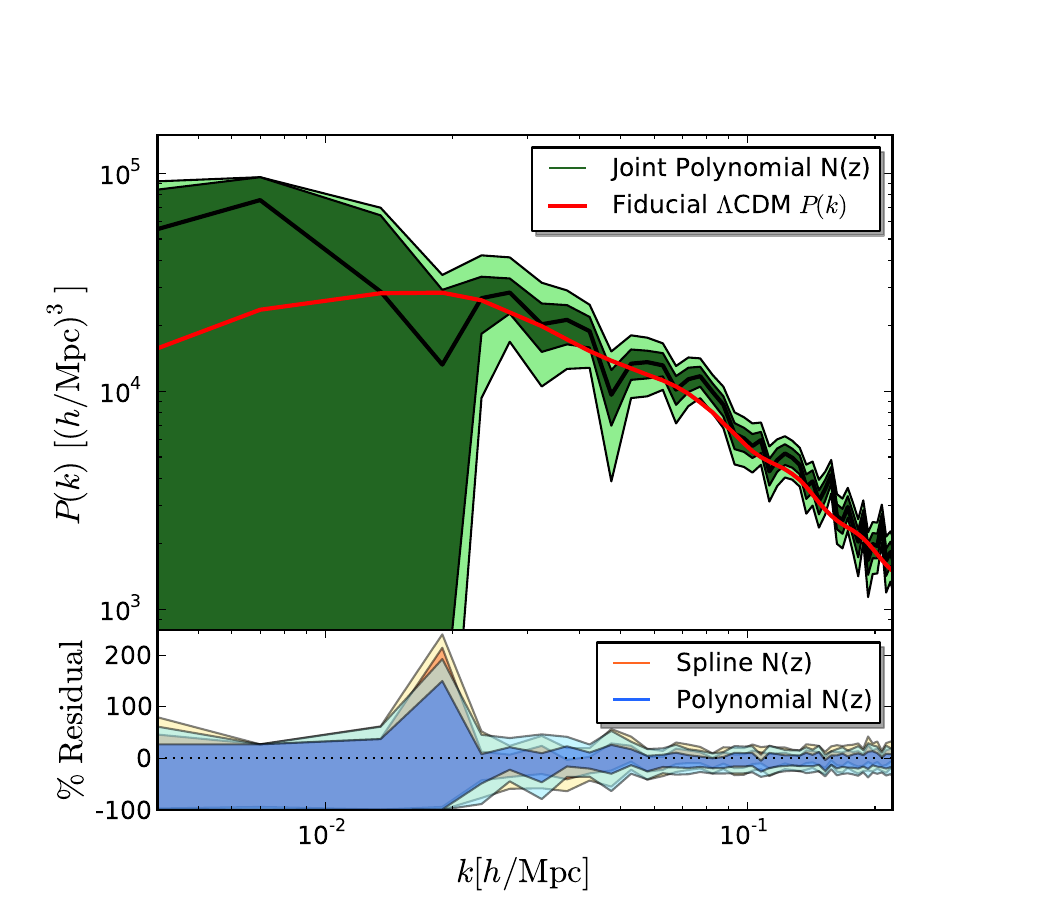}
\caption[deconvolution]{The deconvolved co-added WiggleZ $P(k)$ using 3 different methods for estimating the WiggleZ radial selection function (a cubic spline or a polynomial fit to each region independently and our preferred method: a polynomial fit jointly to the NGP and SGP fields) .  Dark shaded regions depict $68\%$ confidence intervals and light shaded regions depict $95\%$ confidence intervals.  The solid black line depicts the maximum likelihood $P(k)$.  The red line depicts a $\Lambda$CDM model incorporating a simple model for redshift-space distortions.  Residuals (in the bottom panel) depict results after subtraction of the joint spline maximum likelihood $P(k)$, illustrating the magnitude of systematic effects associated with uncertainties in the WiggleZ radial selection function.\label{fig-Pk_deconvolved}} 
\end{figure}

The WiggleZ survey was conducted within seven separate survey regions, each with different selection functions due to varying optical selection, region boundary geometries, observing conditions and dust attenuations.  While much of our analysis in later sections will involve direct fits applied jointly to the data of these regions, we seek here to generate a single deconvolved power spectrum of the full survey sample, with the effects of differing selection functions removed, in order to present the data contained in the survey on its largest scales.

To achieve this we use a Monte Carlo Markov Chain (MCMC) approach.  Using the Metropolis-Hastings algorithm, we generate random sets of propositions for the deconvolved power spectra we seek, convolve each set seven times using the window function of each region and compute the joint likelihood of these convolved proposition sets against the observed power spectrum of each region.  We compute this likelihood using the full information of each region's covariance matrix.  We optimise this calculation by drawing propositions from a rotated covariance matrix constructed from a short secondary burn-in period following a primary burn-in designed to erase the memory of the starting point of our calculation.  We use $10^5$ iterations for each burn-in phase and $2{\times}10^6$ propositions for the final integration used to determine the posterior distribution of our 50 observed, deconvolved power spectrum bins.  All chains have been inspected to ensure that they are well mixed.  To increase the efficiency of our calculation, we use the matrix multiplication approach presented in \citet[][Eqn. 17]{Blake:2010p74} to convolve our $P(k)$ proposition sets with each region's selection function.  We have verified that this approach remains equivalent to a full 3D convolution on all scales utilised for this current analysis.

The results of this calculation are presented in Figure \ref{fig-Pk_deconvolved}.  In this figure we see that at scales $k{<}0.03$ [Mpc/$h$], the deconvolved coadded WiggleZ power spectrum places useful constraints on the maximum power permitted on scales at-and-beyond the turnover, but places little-or-no constraint on the minimum permitted power.  For comparison, we also show (in red) a \LCDM\ power spectrum in our fiducial cosmology with a simple model for redshift-space effects.  This is obtained by taking a biased non-linear \citep[][\ie\ halofit]{Smith:2003p1759} power spectrum from the Boltzmann code CAMB \citep{Lewis:2000} and applying the redshift-space distortion model of \citet{Kaiser:1987p1760} with a Lorentzian damping term \citep[see Equation 9 of][]{Blake:2011b}.  The parameter values of the galaxy bias ($b^2=1.18$), redshift-space distortion parameter ($\beta{=}f{/}b{=}0.69$) and the variable damping term ($\sigma_{\rm v}{=}300$ km/s) needed for this model were obtained from fits to the 2D power spectrum obtained from the WiggleZ NGP fields.  Figure \ref{fig-Pk_deconvolved} illustrates that this model provides an excellent fit to the WiggleZ power spectrum over the full range of scales presented.

\section{Analysis}\label{sec-analysis}

In this section, we will present our extraction of turnover information from the WiggleZ power spectrum presented in Section \ref{sec-observations} as well as the distance and cosmology constraints obtainable from this measurement.

\subsection{Measurement of the WiggleZ Turnover Scale}

\noindent To extract information about the turnover from the observed WiggleZ power spectrum, we have followed the method of \citet{Blake:2005p80}.  This approach is model-independent in the sense that it does not take cosmological parameters or model power spectra from Boltzmann codes as inputs.  Specifically, we fit (using the same MCMC machinery and chain lengths descibed in Section \ref{sec-Pk_deconvolved}) the following model, convolved with the WiggleZ selection functions and jointly fit to all 7 WiggleZ regions:
\begin{align} \label{eqn-turnover_model}
\log_{10} P(k)= &
      \begin{cases}
         \log_{10} P_{\rm 0} \left( 1- \alpha x^2 \right) & \text{if $k < k_{\rm 0}$,} \\
         \log_{10} P_{\rm 0} \left( 1- \beta   x^2 \right) & \text{if $k \ge k_{\rm 0}$}
      \end{cases} \\
\text{where } x= & \left( \frac{\ln k-\ln k_{\rm 0}}{\ln k_{\rm 0}} \right) 
\end{align}

\noindent We marginalise over $P_{\rm 0}$, since its interpretation is complicated by degenerate normalising parameters such as $\sigma_{\rm 8}$ and the bias of WiggleZ galaxies, and over $\beta$ since its interpretation is complicated by all the cosmological parameters, redshift-space and scale-dependent bias effects which influence the overall shape of $P(k)$ on scales smaller than the turnover.  Hence, we focus here on just two parameters:  $k_{\rm 0}$ and $\alpha$, which carry information about the position of the peak in $P(k)$ and whether we have detected a drop in power at scales larger than this peak (this is the case if $\alpha>0$).  All cosmology constraints derived from our measurement of the turnover will be based purely on our measurement of $k_{\rm 0}$.  For our fiducial \LCDM\ cosmology, the expected value is $k_{\rm 0}{=}0.016$ [$h$/Mpc].

\begin{figure}
\includegraphics[width=84mm]{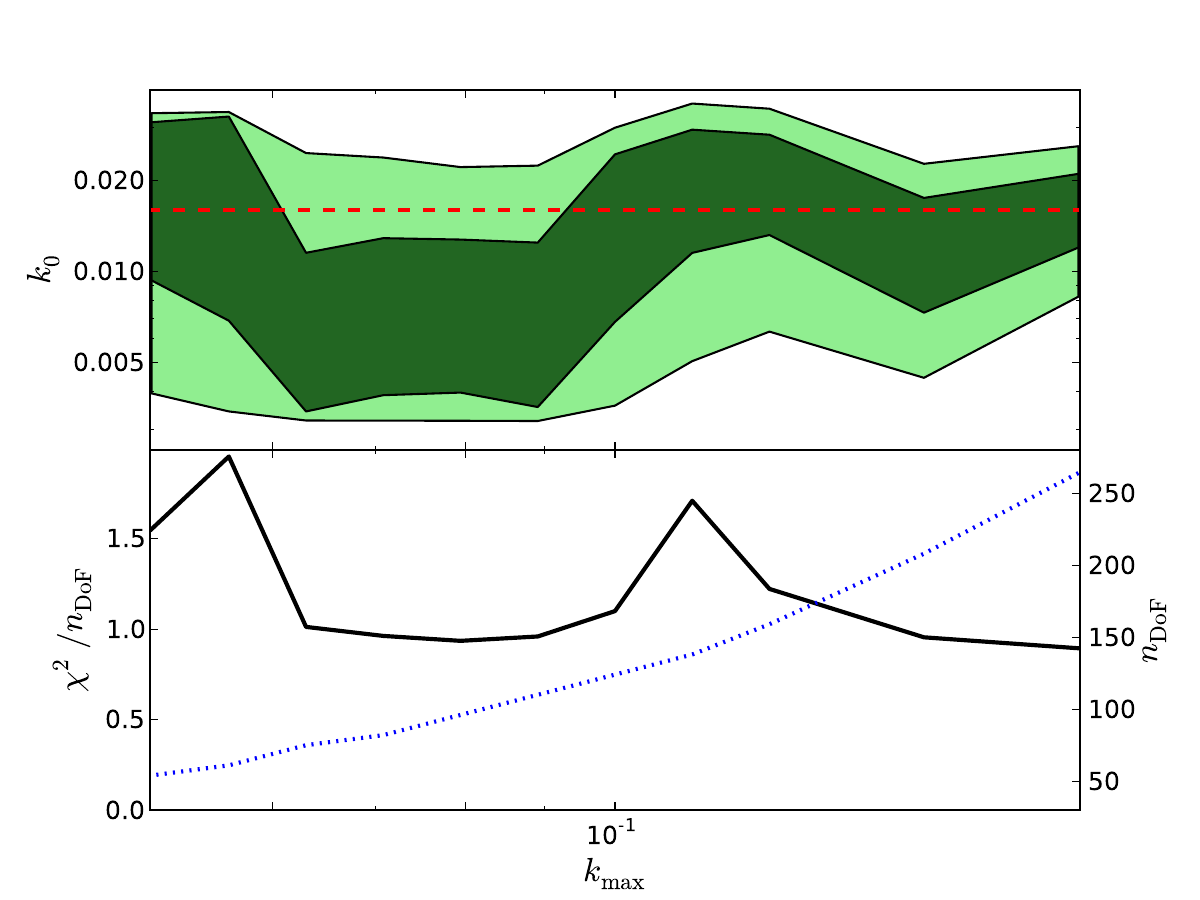}
\caption[ko, alpha and chi2 vs. kmax]{Results of fitting our logarithmic parabola turnover model (see Eqn. \ref{eqn-turnover_model}) to the observed WiggleZ $P(k)$ as a function of the maximum range in $k$ used for the fit ($k_{\rm max}$).  Top panel: dark shaded regions depict $68\%$ confidence intervals and light shaded regions depict $95\%$ confidence intervals.  The red dashed line depicts the theoretical expectation for a standard $\Lambda$CDM model.  Bottom panel: the black line depicts the reduced-$\chi^2$ statistic and the dotted blue line depicts the number of degrees of freedom ($n_{\rm DoF}$).  We see from this that our measured turnover scale is stable and yields an acceptable fit for $k_{\rm max}{\gtrsim}0.1$.\label{fig-kmax_plot}} 
\end{figure}

Since the real power spectrum has a changing logarithmic slope with $k$ at scales smaller than the turnover, we need to carefully consider the maximum of the range of scales over which we perform our fit (denoted $k_{\rm max}$) to avoid systematic biases.  We have performed our fit using a range of $k_{\rm max}$ and present the results in Figure \ref{fig-kmax_plot}.  Here we see that the results of our fit for $k_{\rm 0}$ are both stable and acceptable (as measured by the reduced-$\chi^2$ statistic) for $k_{\rm max}{\gtrsim}0.1$.  However, non-linear effects which alter the logarithmic slope of the power spectrum are expected to become significant for $k{>}0.2$ [$h$/Mpc] and so we limit our fit with $k_{\rm max}{=}0.2$ [$h$/Mpc] for the remainder of our analysis.

\begin{figure}
\includegraphics[width=84mm]{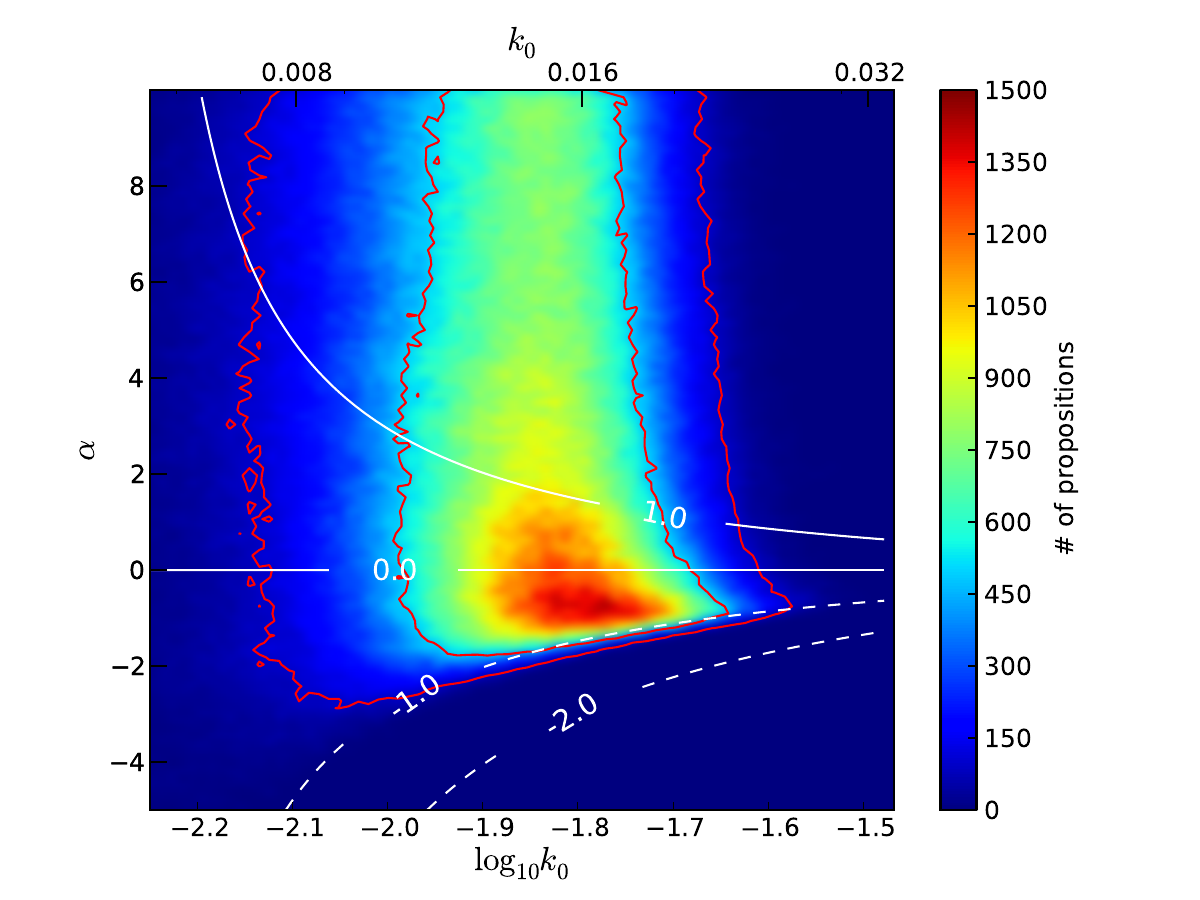}
\caption[PDF of alpha vs. ko]{The 2D marginalised posterior distribution function in the $\alpha$-$k_{\rm 0}$ plane from fitting our logarithmic parabola turnover model (see Eqn. \ref{eqn-turnover_model}) to the observed WiggleZ $P(k)$.  Red contours depict the $68\%$ and $95\%$ confidence regions.  The white contours depict the loci of several values of $n$ for the model $P(k{\leq}k_{\rm 0})\propto k^n$.  From this plot we see that the WiggleZ $P(k)$ has an inflection at $k_{\rm 0}{\sim}0.016$ [$h$/Mpc] with $n{>}{-1}$, distinctly different from the asymptotic value for $P(k{\gg}k_{\rm 0})$ of $n{=}{-}3$.  Our fit easily accommodates the theoretical expectation of $k_{\rm 0}{=}0.016$ [$h$/Mpc] and $n{=}{1}$ for a standard \LCDM\ model.   \label{fig-pdf_alpha_ko}} 
\end{figure}

The final marginalised results for $k_{\rm 0}$ from our fit are $k_{\rm 0}{=}0.0160^{+0.0041}_{-0.0035}$ ($68$\% confidence); $^{+0.0075}_{-0.0073}$ ($95$\% confidence).  Our results are not well constrained for $\alpha$ where we measure lower limits of $\alpha{>}{-1.32}$ ($95$\%) with a hard prior $\alpha{<}10$.  In Figure \ref{fig-pdf_alpha_ko} we present the 2D-marginalized posterior distribution function of our fit to the WiggleZ power spectrum in the $\alpha{-}k_{\rm 0}$ plane.   We can see from this figure that our analysis constrains $k_{\rm 0}$ to lie near the theoretical value of $k_{\rm 0}{\sim}0.016$ [$h$/Mpc].  Furthermore, $\alpha$ is constrained to $\alpha{\gtrsim}{-}1$ indicating that $P(k)$ is either an increasing function of $k$ or nearly constant at scales larger than the turnover.  The long tail to positive values of $\alpha$ reflects the lack of constraint WiggleZ places on the minimum power allowed on the largest scales of the survey.  

Negative values of $\alpha$ are marginally allowed by our fit, permitting a power spectrum which does not formally exhibit a turnover.  To aid our interpretation of this result we have overlayed onto Figure \ref{fig-pdf_alpha_ko} (with white contours) a model which presents the logarithmic slope (denoted $n$) for $P(k{<}k_{\rm 0}){\propto}k^n$ implied by Equation \ref{eqn-turnover_model} between $k_{\rm 0}$ and our largest-scale bin at $k{=}0.005$ [$h$/Mpc].  This model incorporates the strong and tight degeneracy
\begin{equation}\label{eqn-Po_ko_degeneracy}
      \log_{10}P_{\rm 0}=-0.67\log_{10}k_{\rm 0}+3.12
\end{equation}

\noindent present in our fit.  From this we can see that WiggleZ constrains the power at scales larger than the turnover to $n{>}{-}1$ at nearly $95\%$ confidence.  The large-$k$ asymptotic value of $n{=}{-}3$ is completely ruled out, meaning that we have certainly detected an inflection in the logarithmic slope of $P(k)$.  With the theoretically expected value of $n{=}1$ easily accommodated by our fit, and a preferred scale for the $P(k)$ inflection of $k_{\rm 0}{\sim}0.016$ [$h$/Mpc], we thus find that our turnover fit is in good agreement with the theoretical expectations of a standard \LCDM\ model. 

\subsection{The WiggleZ Turnover Distance Measurement}

The scale of the turnover roughly corresponds to the scale of the horizon at the epoch of matter-radiation equality ($k_{\rm H}$) given by \citep[see][]{Prada:2011p1771}
\begin{align} \label{eqn-turnover_model}
   k_{\rm H}= & {\left(4-2\sqrt{2}\right)}{r_{\rm H}^{-1}} \\
   \text{where } r_H = & c \int_0^{\left(1+z_{\rm eq}\right)^{-1}}{\frac{\,da}{a^2H(a)}} 
\end{align}
\noindent with $H(a)$ being the Hubble expansion rate given by
\begin{equation} \label{eqn-Hubble_rate}
   H^2(a)= H^2_{\rm 0} \left[ \Omega_{\rm r}a^{-4}{+}\Omega_{\rm M}a^{-3}{+}\Omega_\Lambda \right]
\end{equation}
\noindent and $H_{\rm 0}$ being the Hubble constant.  This yields a horizon scale of $r_{\rm H}{=}117.9$ [Mpc] corresponding to $k_{\rm H}{=}0.014$ [$h$/Mpc] for the WMAP result of $z_{\rm eq}=3145$ \citep{Komatsu:2011p1737} in our fiducial \LCDM\ cosmology.  However, the actual precise position of the peak of $P(k)$ can differ significantly from this value.  Indeed, using the Boltzmann code CAMB \citep{Lewis:2000} we find that the turnover scale in our fiducial cosmology is $k_{\rm 0,fid}{=}0.0159$ [$h$/Mpc].

As discussed in Section \ref{sec-radial_selection}, the largest scales of the power spectrum utilised in this work are dominated by radial modes.  However, for most of the scales involved in our turnover fit, this is not the case.  For this reason, we choose to convert our measurement of the turnover scale into a distance measurement following the approach introduced by \citet{Eisenstein:2005p1275} for BAO studies. Through this approach, our turnover measurement constrains the dilation measure $D_{\rm V}$ defined as
\begin{equation}\label{eqn-dilation_measure}
D_{\rm V}(z)=\left[ (1+z)^2D_{\rm A}^2(z)\frac{cz}{H(z)} \right]^{1/3}
\end{equation}
\noindent where $c$ is the speed of light, $D_{\rm A}(z)$ is the angular diameter distance at redshift $z$ and $H(z)$ is given by Equation \ref{eqn-Hubble_rate} with $a=\left( 1+z \right)^{-1}$.  We will obtain this quantity using the assumption that distances scale proportionally by the same ``stretch factor'' ($\tilde{\alpha}$) under small perturbations from a fiducial cosmology.  Given the turnover scale in our fiducial cosmology, we obtain this stretch factor from
\begin{equation} \label{eqn-stretch_factor}
   \tilde{\alpha}{=}k_{\rm 0,fid}/k_{\rm 0}
\end{equation}

\noindent We will express this dilation measure in a dimensionless form using units of the $z{=}3145$ horizon size ($r_{\rm H}$) as $d_{\rm t}{=}D_{\rm V}{/}r_{\rm H}$.  We choose this scale instead of the turnover scale from CAMB to render the result less model-dependant.

For this method we require an effective redshift for our measurement, which we compute by determining the effective redshift at $k{=}0.015$ [$h$/Mpc] for the power spectrum used in this analysis.  This is achieved using Equation 13 of \citet{Blake:2011a} from which we obtain $z_{\rm eff}{=}0.62$; quite similar to the effective redshift of other WiggleZ studies.

The dilation measure at $z_{\rm eff}$ in our fiducial cosmology is $d_{\rm t,fid}(z_{eff}{=}0.62){=}18.41$.  For our measurement of $k_{\rm 0}{=}0.0160^{+0.0035}_{-0.0041}$ ($68$\%); $^{+0.0073}_{-0.0075}$ ($95$\%) we obtain a stretch factor $\tilde{\alpha}{=}0.99^{+0.34}_{-0.18}$ ($68$\%); $^{+1.07}_{-0.37}$ ($95$\%).  Scaling with respect to our fiducial cosmology using this stretch factor, we thus obtain a model independent distance of $d_t(z_{\rm eff}{=}0.62){=}18.3^{+6.3}_{-3.3}$ ($68$\%); $^{+19.7}_{-6.7}$ ($95$\%).  Assuming the fiducial turnover scale given above, this corresponds to $D_{\rm V}(z_{\rm eff}{=}0.62){=}2156^{+743}_{-387}$ ($68$\%); $^{+2324}_{-792}$ ($95$\%) [Mpc].

\subsection{Cosmology Constraints from the WiggleZ Turnover Measurement}

To calibrate the scale of the turnover as a function of cosmology we ran CAMB to produce a series of matter power spectra, fixing all parameters to our fiducial cosmology but allowing $\Omega_{\rm M} h^2$ to vary for several choices of $N_{\rm eff}$.  The results are presented in Figure \ref{fig-Neff_OmM} where we illustrate our method of converting measurements of $k_{\rm 0}$ into cosmological constraints.

\begin{figure}
\includegraphics[width=84mm]{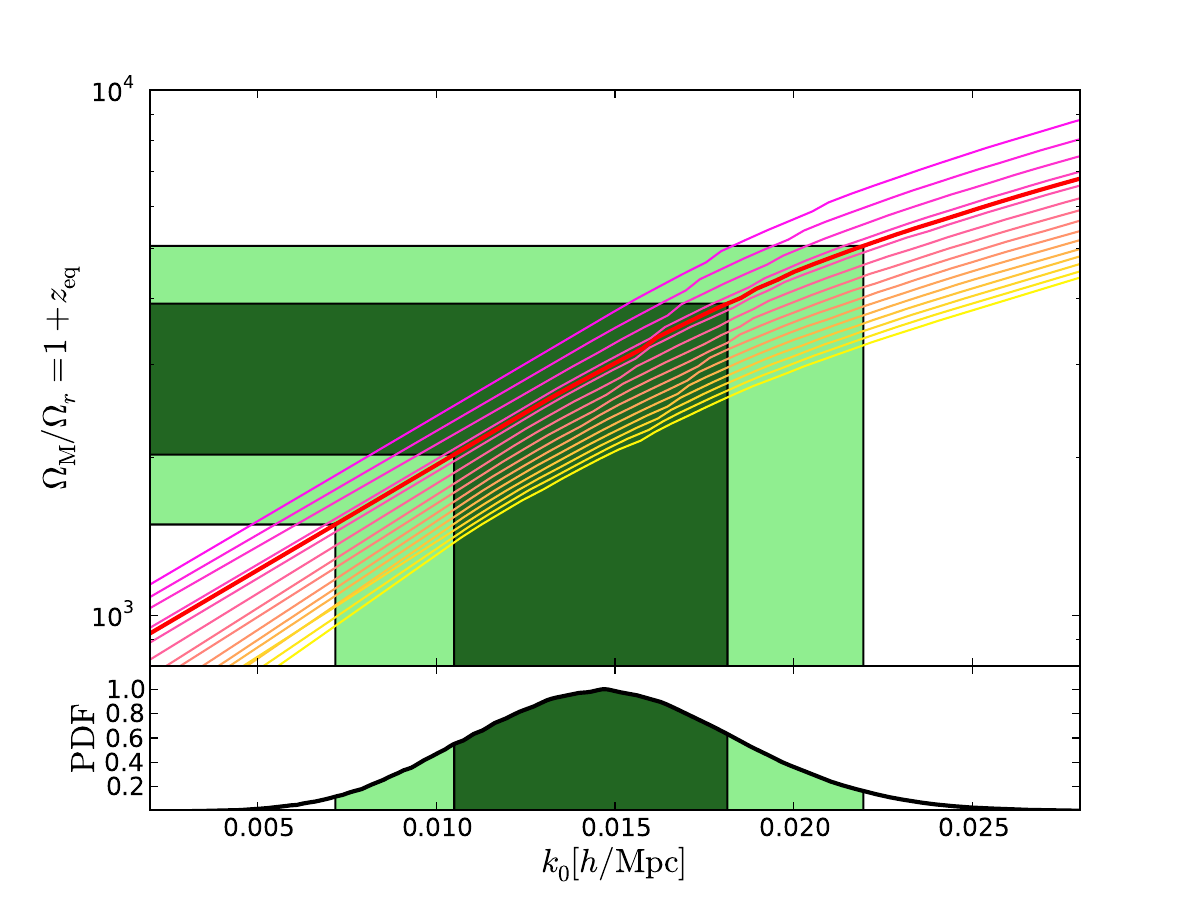}
\caption[ko vs. OmOrel]{Bottom panel: The 1D marginalised posterior distribution function (PDF) for $k_{\rm 0}$ from fitting our logarithmic parabola turnover model (see Eqn. \ref{eqn-turnover_model}) to the observed WiggleZ $P(k)$.  Dark shaded regions depict $68\%$ confidence intervals and light shaded regions depict $95\%$ confidence intervals.  Top panel: an illustration of how our constraint on $k_{\rm 0}$ maps to a constraint on $\Omega_{\rm M}/\Omega_{\rm r}{=}1+z_{\rm eq}$.  Lines depict the dependance of $1+z_{\rm eq}$ on $k_{\rm 0}$, calibrated using CAMB, for several values of $N_{\rm eff}$ (increasing from $N_{\rm eff}{=}0$ in magenta to $N_{\rm eff}{=}10$ in yellow, in steps of 0.67; $\Omega_{\rm M} h^2$ increases from left to right).  We highlight the relation for $N_{\rm eff}{=}3$ in red as well as the constraint on $1+z_{\rm eq}$ determined from our measurement of $k_{\rm 0}$ for this case.\label{fig-ko_OmOrel}} 
\end{figure}

The position of the turnover depends on $\Omega_{\rm M} h^2$ and $N_{\rm eff}$ in the following way  \citep[see][]{Komatsu:2009}.  First, through the dependance of the total energy density of relativistic material on $N_{\rm eff}$ given by
\begin{align}\label{eqn-Omega_rel}
\Omega_{\rm r}= & \Omega_{\gamma}\left[1+0.2271N_{\rm eff} \right] \\
\text{where } \Omega_{\gamma}= & \frac{8 \pi G}{3H_{\rm 0}^2}\frac{4 \sigma_{\rm B}T^4_{\rm CMB}}{c^3}
\end{align}

\noindent with $\Omega_{\gamma}$ being the energy density of photons\footnotemark; $G$ Newton's gravitational constant; $\sigma_{\rm B}$ the Stefan-Boltzmann constant; and $T_{\rm CMB}$ the temperature of the CMB, and second, through the dependance of the redshift of matter-radiation equality on $\Omega_{\rm M}$ and $\Omega_{\rm r}$ given by
\begin{equation}\label{eqn-z_eq}
      1+z_{\rm eq}=\frac{\Omega_{\rm M}}{\Omega_{\rm r}}=\frac{\Omega_{\rm M}h^2}{\Omega_{\gamma}h^2}\frac{1}{1+0.2271N_{\rm eff}}
\end{equation}

\footnotetext{Note that due to the precision with which the CMB temperature is measured \citep[$T_{\rm CMB}{=}2.72548{\pm}0.00057$K,][]{Fixsen:2009p1752}, the photon energy density is precisely known to be $\Omega_{\gamma}h^2{=}2.47274{\times}10^{-5}$.}

\noindent From this we can convert our constraints on the turnover scale into constraints on $\Omega_{\rm M} h^2$, $N_{\rm eff}$ and subsequently $z_{\rm eq}$ as follows.  In Figure \ref{fig-ko_OmOrel} we plot the ratio $\Omega_{\rm M}/\Omega_{\rm r}$ as a function of the turnover scale ($k_{\rm 0}$) measured from our CAMB power spectra for a range of $N_{\rm eff}$ between $0$ and $10$.  Each line represents results for a wide range of values for $\Omega_{\rm M} h^2$, which increases with increasing $k_{\rm 0}$.  Projecting an observed turnover scale onto this sequence of curves allows us to map any value of $k_{\rm 0}$ to a unique set of values for $\Omega_{\rm M} h^2$ and $\Omega_{\rm M}/\Omega_{\rm r}{=}1{+}z_{\rm eq}$ as a function of $N_{\rm eff}$.

To see the form of the resulting constraint in the $N_{\rm eff}{-}\Omega_{\rm M} h^2$ plane we can rearrange Equation \ref{eqn-z_eq} to get
\begin{equation}\label{eqn-Neff_degeneracy}
      N_{\rm eff}=4.403 \left[ \left( \frac{\Omega_{\rm M}}{\Omega_{\rm r}} \right)^{-1}{\left( \Omega_{\gamma}h^2 \right)}^{-1} \left(\Omega_{\rm M}h^2 \right) -1 \right]
\end{equation}

\noindent where the first bracketed factor is constrained from the turnover measurement and the second bracketed factor is known precisely from the CMB.  Hence, we expect the scale of the turnover to place a roughly linear degenerate constraint in this plane, since our one measurement can not place closed constraints on two values.

\begin{figure}
\includegraphics[width=84mm]{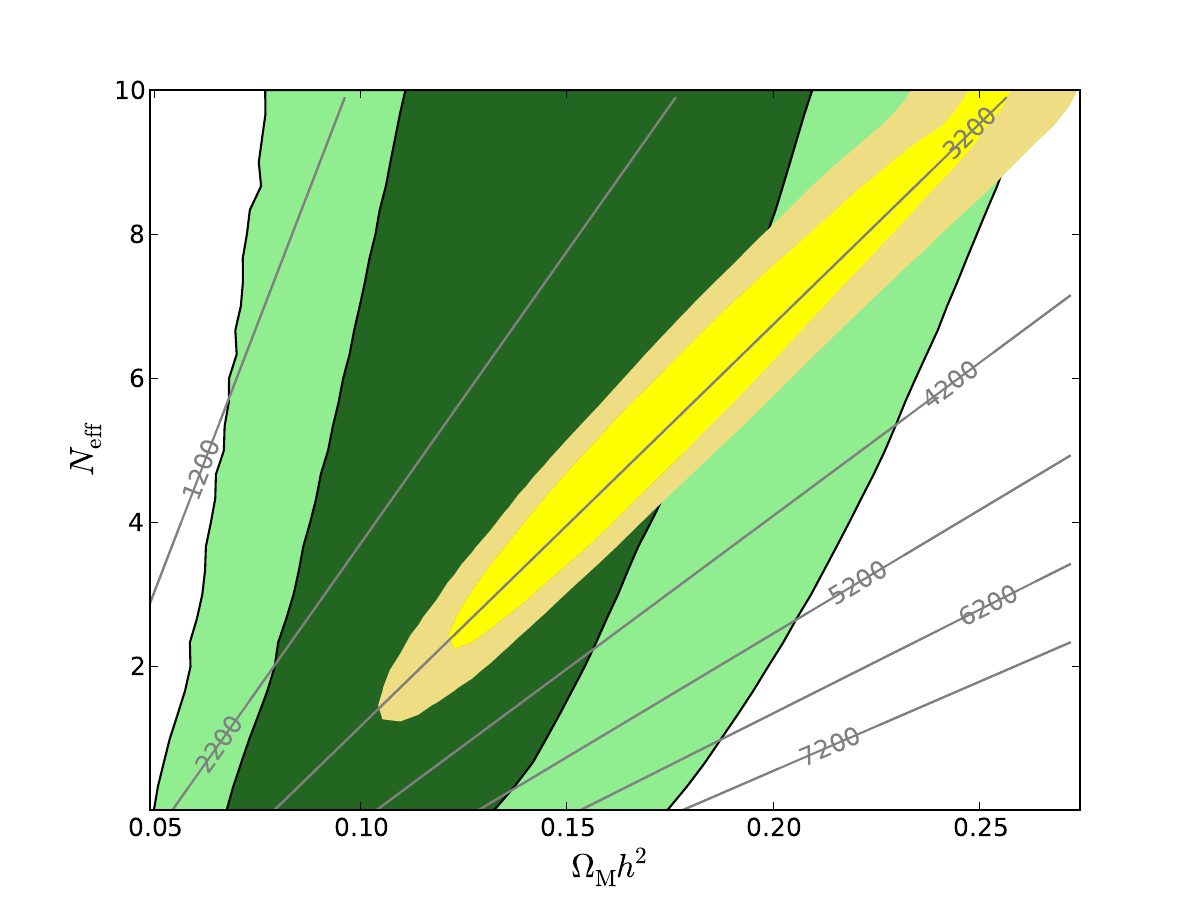}
\caption[Neff vs. Omega M]{The constraints in the $N_{\rm eff}$-$\Omega_{\rm M} h^2$ plane derived from the WiggleZ measurement of the turnover scale (in green) compared to CMB CosmoMC constraints from WMAP (in yellow).  Dark shaded regions depict $68\%$ confidence intervals and light shaded regions depict $95\%$ confidence intervals.  Grey contours depict lines of constant $\Omega_{\rm M}/\Omega_{\rm r}{=}1{+}z_{\rm eq}$. \label{fig-Neff_OmM}} 
\end{figure}

We also show on Figure \ref{fig-Neff_OmM} the WMAP CMB constraint in the $N_{\rm eff}{-}\Omega_{\rm M} h^2$ plane. We made use of the \LCDM+$N_{\rm eff}$ WMAP 7-year chains, downloaded from the WMAP LAMBDA Data products site\footnote{http://lambda.gsfc.nasa.gov/product/map/current/}, and analysed them using the GetDist package that comes as part of CosmoMC \citep{Lewis:2002p1743}.

Immediately obvious from this plot is the degeneracy in the WiggleZ constraints expressed by Equation \ref{eqn-Neff_degeneracy}.  The WMAP constraints also exhibit a strong degeneracy, although oriented differently and  along  the line $1{+}z_{\rm eq}{\sim}3200$.  While this degeneracy does not close for the WiggleZ constraint, it does for the WMAP constraint, demanding $N_{\rm eff}{\gtrsim}2$.  While we do not perform a rigorous joint fit of these measurements, it is clear that the WiggleZ turnover constraint should help to close the CMB constraints at an upper limit $N_{\rm eff}{\lsim}9$.  

While our measurement of the WiggleZ turnover scale clearly provides little constraint on $N_{\rm eff}$ (even with a strong prior on $\Omega_{\rm M} h^2$), a constraint on $\Omega_{\rm M} h^2$ can be derived given a prior on $N_{\rm eff}$.  Assuming the value $N_{\rm eff}{=}3$, we obtain a WiggleZ turnover constraint of $\Omega_{\rm M}h^2{=}0.136^{+0.026}_{-0.052}$ ($68$\%); $^{+0.073}_{-0.074}$ ($95$\%).  As illustrated by Equation \ref{eqn-z_eq}, given a strong prior on $N_{\rm eff}$, the fractional constraint on $\Omega_{\rm M} h^2$ maps almost identically to a fractional constraint on $z_{\rm eq}$.  Thus, we find that the WiggleZ turnover provides the constraint $z_{\rm eq}{=}3274^{+631}_{-1259}$ ($68$\%); $^{+1757}_{-1791}$ ($95$\%).  All results are unchanged if the expected value $N_{\rm eff}=3.046$ is taken for the assumed prior instead.

\section{Forecasts for future surveys}\label{sec-forecasts}
From these results, the question naturally arises: what sort of constraint on $N_{\rm eff}$, $\Omega_{\rm M} h^2$ and $z_{\rm eq}$ would be possible from a larger survey such as BOSS or Euclid, encompassing a larger volume and hence, providing a stronger turnover constraint?

In this section we examine the prospects of using the turnover position in current and future surveys for cosmology constraints.  The key questions we seek to answer are: how effective will turnover constraints on $N_{\rm eff}$ and $\Omega_{\rm M} h^2$ be from BOSS and Euclid and how big does a survey need to be for these constraints to be competitive with those from CMB observations?

We will find that the required volumes are large but it is important to note that photometric redshift surveys for which imaging systematics are under good control are just as effective for this science as spectroscopic redshift surveys.  This is because the distance uncertainties associated with photometric redshift errors will be small compared to the scales of the turnover \citep{Blake:2005p80}.

\subsection{Constructing mock survey constraints}

We have repeated the analysis presented in Section \ref{sec-analysis} on a series of mock power spectra generated using CAMB in our fiducial cosmology and given ranges in $k$ and noise properties designed to represent future surveys.  We ignore the complicating issues of survey selection and geometry and assume that the scales relevant to this analysis remain volume limited in their precision.  Under these conditions, survey volume (denoted as $V$) is the only survey parameter relevant to our calculation.  In all cases, we assume that the largest scale probed is $L_{\rm max}=\sqrt[3]{V}$ and choose a power spectrum binning given by $\Delta k=2\pi/L_{\rm max}$.  In each case we combine 20 chains for the analysis of each forecasted survey volume, each with differing random seeds for the noise generation process.

For these calculations we have changed the maximum value of $k$ over which we perform our turnover fits to $k_{\rm max}{=}0.04$ [$h$/Mpc].  This was done to avoid a systematic bias introduced by the BAO features of the power spectrum which becomes significant when volumes exceed those of WiggleZ.  This bias is driven by a degeneracy between $\beta$ and $k_0$ and a dependance of $\beta$ on $k_{\rm max}$.

\begin{figure}
\includegraphics[width=84mm]{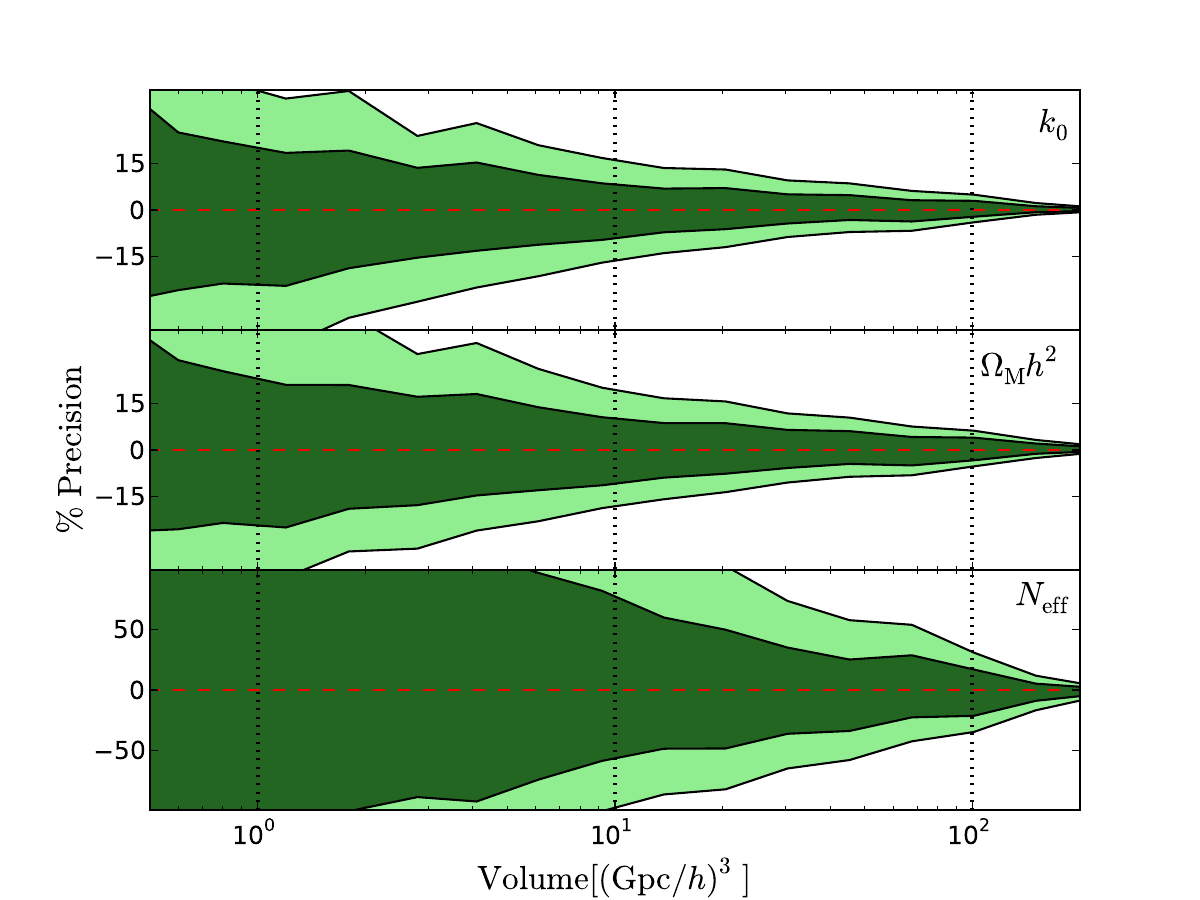}
\caption[Cosmology forecasts.]{The precision of turnover scale ($k_{\rm 0}$) measurements and subsequent constraints on $\Omega_{\rm M} h^2$ and $N_{\rm eff}$ as a function of survey volume.  Vertical dotted lines indicate the approximate volume of WiggleZ, BOSS and Euclid in order of increasing volume.  For the $\Omega_{\rm M}h^2$ constraints an $N_{\rm eff}{=}3$ prior is assumed and for the $N_{\rm eff}$ constraints an $\Omega_{\rm M}h^2=0.135$ prior is assumed (although results are insensitive to these choices).  Dark shaded regions depict $68\%$ confidence intervals and light shaded regions depict $95\%$ confidence intervals.\label{fig-forecast_cosmology}} 
\end{figure}

We also slightly change the functional form of our model power spectrum for this calculation.  With sufficiently large volume, the primordial power spectrum should begin to emerge in a survey's largest modes.  The primordial power spectrum is expected to be a power law (or nearly-so), which differs substantially from the simple asymmetric logarithmic parabola model of Equation \ref{eqn-turnover_model}.
To capture this behaviour in our forecasts for surveys with very large volumes, we instead use for our power spectrum model an asymmetric logarithmic hyperbola on scales larger than the turnover.  Equation \ref{eqn-turnover_model} then becomes:
   \begin{equation} \label{eqn-turnover_forecast}
   \log_{10} P(k)= 
      \begin{cases}
         \log_{10} P_{\rm 0}{+}p{-}\sqrt{\alpha^2 x^2{+}p^2} & \text{if $k < k_{\rm 0}$,} \\
         \log_{10} P_{\rm 0} \left( 1- \beta   x^2 \right) & \text{if $k \ge k_{\rm 0}$.}
      \end{cases} 
   \end{equation}
\noindent This model adds an additional degree of freedom $(p)$ which describes how quickly $P(k)$ asymptotes to a power law at $k{<}k_{\rm 0}$.  We have verified that using this model on the observed WiggleZ $P(k)$ and on model power spectra of comparable volume generates consistent results for the turnover scale.

\subsection{Forecast results}

The results of our survey forecasts are shown in Figure \ref{fig-forecast_cosmology} where we present, as a function of survey volume, the precision of the turnover scale measurement and of the resulting constraints on $\Omega_{\rm M} h^2$ and $N_{\rm eff}$.  For the $\Omega_{\rm M} h^2$ constraints an $N_{\rm eff}{=}3$ prior is assumed and for the $N_{\rm eff}$ constraints an $\Omega_{\rm M}h^2{=}0.135$ prior is assumed.  Results are insensitive to these choices however, particularly for large volumes.

In this figure we highlight the results for survey volumes approximately equal to those of WiggleZ, BOSS and Euclid representing a series covering the range of our calculations in logarithmically equal steps in volume.  The BOSS survey \citep{Eisenstein:2011p1756} will map $10000$ square degrees, collecting spectroscopic redshifts for luminous red galaxies out to $z{\sim}0.7$.  With completion planned for 2014, BOSS represents the largest survey for which near-term results will be possible.  The Euclid satellite \citep{Laureijs:2011p1755} will conduct a wide-field ($15000$ square degree) extragalactic survey covering redshifts out to $z{\sim}2$.  With a target launch date of 2017-2018, the anticipated specifications of this survey represent a rough limit to our capabilities for conducting the science discussed here at the turn of the next decade.

\begin{figure}
\includegraphics[width=84mm]{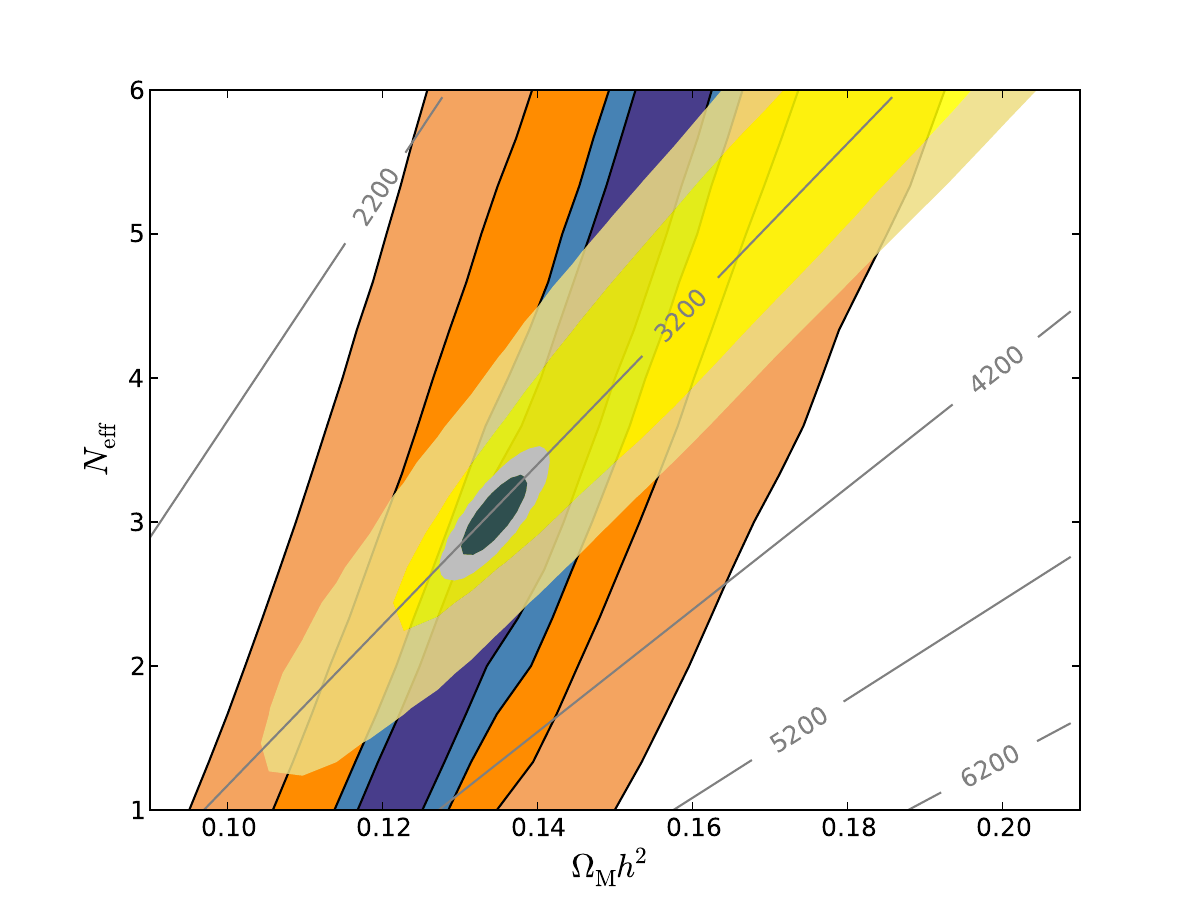}
\caption[Neff vs. Omega M forecast]{The constraints in the $N_{\rm eff}$-$\Omega_{\rm M}h^2$ plane derived from our forecasts for the BOSS (in orange) and Euclid (in blue) measurements of the turnover scale.  These are compared to the CMB constraint from WMAP (in yellow) and our Planck forecast (in grey).  Dark shaded regions depict $68\%$ confidence intervals and light shaded regions depict $95\%$ confidence intervals.  Grey contours depict lines of constant $\Omega_{\rm M}/\Omega_{\rm r}{=}1{+}z_{eq}$.  \label{fig-Neff_OmM_forecast}} 
\end{figure}

In Figure \ref{fig-forecast_cosmology} we see that results for $V{\sim}1$ [(Gpc/h)$^3$] are roughly consistent with the WiggleZ constraints  (although slightly better, presumably due to the lack of covariance in our forecast model) presented in Figures \ref{fig-pdf_alpha_ko}, \ref{fig-ko_OmOrel} and \ref{fig-Neff_OmM}: a $k_{\rm 0}$ precision of ${\pm}20$\% ($68$\%); ${\pm}40$\% ($95$\%) with little resulting constraint on $N_{\rm eff}{\lsim}10$.  The $24$\% precision of the constraint on $\Omega_{\rm M}h^2$ (and thus, $z_{\rm eq}$) for this case maps almost directly to the precision of the constraint on $k_{\rm 0}$.

As volume approaches that of BOSS at $V{\sim}10$ [(Gpc/h)$^3$], the nature of these constraints changes significantly.  With a sizeable increase in the number and accuracy of measured modes and scales beyond the turnover, the constraints on $k_{\rm 0}$ improve dramatically to a precision of ${\pm}9$\% ($68$\%); ${\pm}16$\% ($95$\%).  This translates to significantly improved constraints on $\Omega_{\rm M} h^2$ and $z_{\rm eq}$ of ${\pm}10$\% ($68$\%); ${\pm}20$\% ($95$\%).  The constraints on $N_{\rm eff}$ start to become interesting at a level of ${}^{+78}_{-56}$\% ($68$\%); ${}^{+152}_{-98}$\% ($95$\%), as well.

Looking past BOSS to much larger future surveys, precision improves more slowly with increasing volume.  However, with $V{\sim}100$ [(Gpc/h)$^3$], the proposed volume of Euclid\footnotemark{} is a dramatic order-of-magnitude increase over BOSS.  As a result, Euclid should ultimately be able to measure $k_{\rm 0}$ to a precision of ${\pm}3$\% ($68$\%); ${\pm}5$\% ($95$\%) with corresponding constraints on $\Omega_{\rm M} h^2$ and $z_{\rm eq}$ of ${\pm}4$\% ($68$\%); ${\pm}6$\% ($95$\%) and $N_{\rm eff}$ of ${}^{+17}_{-21}$\% ($68$\%); ${}^{+32}_{-35}$\% ($95$\%).

\footnotetext{From the Euclid Science Requirements Document, ESA Science document reference number DEM-SA-Dc-00001.}

We contrast these forecasts with the results anticipated from Planck.  For our Planck constraints we generated simulated Planck data following the procedure used in \citet{Perotto:2006p1757} using the specifications for the HFI bolometers given in the Planck Blue Book \citep{ThePlanckCollaboration:2006p1758}.  We then used CosmoMC to generate chains, assuming a \LCDM+$N_{\rm eff}$ model.  

In Figure \ref{fig-Neff_OmM_forecast} we illustrate our forecasted constraints for BOSS and Euclid in the $N_{\rm eff}$-$\Omega_{\rm M}h^2$ plane, compared to the CMB constraints of WMAP and our forecasts for the Planck Surveyor.  We can see from this figure that BOSS should make important contributions to the present WMAP constraint, closing the upper limit of the $68$\% confidence contours at $N_{\rm eff}{\lsim}6$.  Looking to the future however, we see the dramatic improvement Planck will make to this measurement.  In detail, we find the constraints from Planck on ($\Omega_{\rm M}h^2$, $N_{\rm eff}$, $z_{\rm eq}$) should be ${\pm}3$, ${\pm}9$, and ${\pm}2$\% ($68$\% confidence); ${\pm}5$, ${\pm}14$, and ${\pm}4$\% ($95$\% confidence) respectively.  While this constraint on $z_{\rm eq}$ is significantly better than what Euclid will provide on its own (even with strong priors), the constraints on $\Omega_{\rm M} h^2$ and $N_{\rm eff}$ are otherwise comparable.


\section{Summary and conclusions}\label{sec-summary}

We have presented an analysis of the WiggleZ Dark Energy Survey, constructing a galaxy power spectrum optimised for studying the largest scales of the survey.  We extract from this the most robust measurement to date of the scale of the turnover in the Universe's matter power spectrum.  From this, we have obtained the first distance measurement and cosmology constraints yet derived from measurements of this feature.  We have also constructed forecasts for the precision of this analysis for future surveys, contrasting the constraints we expect to obtain from turnover measurements in future redshift surveys to complementary constraints from the published observations of WMAP and our forecasts for the Planck Surveyor.

\begin{table*}
\caption[Measurement and forecast results]{A summary of our WiggleZ turnover-derived cosmology measurements and of our forecasts for the precisions of turnover-derived and CMB results in future/ongoing surveys. Unbracketed quantities are $68$\% confidence results and bracketed quantities are $95$\% confidence results.\label{table-results}}
\begin{tabular}{lcccccccc}
\hline
Survey                            &
\multicolumn{2}{c}{$k_{\rm 0}$} &
\multicolumn{2}{c}{$\Omega_{\rm M} h^2$}   &
\multicolumn{2}{c}{$z_{\rm eq}$} &
\multicolumn{2}{c}{$N_{\rm eff}$}  \\
\hline
WiggleZ Measurement		&  $0.0160^{+0.0035}_{-0.0041}$	& (${}^{+0.0073}_{-0.0075}$)	& $0.136^{+0.026}_{-0.052}$ & (${}^{+0.073}_{-0.074}$) & $3274^{+632}_{-1260}$ & ($^{+1757}_{-1791}$) & -- & --  \\
\hline
BOSS Forecast	 Precision		&  ${\pm}9$\% & (${\pm}16$\%) & ${\pm}10$\% & (${\pm}20$\%) & ${\pm}10$\% & (${\pm}20$\%) & ${}^{+78}_{-56}$\% & (${}^{+152}_{-98}$\%)  \\
Euclid Forecast	 Precision		&  ${\pm}3$\% & (${\pm}5$\%) & ${\pm}4\%$ & (${\pm}6\%$) & ${\pm}4$\% & (${\pm}6$\%) & ${}^{+17}_{-21}$\% & (${}^{+32}_{-35}$\%)  \\
Planck Forecast Precision	&  --			& --			& ${\pm}3\%$ & (${\pm}5\%$) & ${\pm}2$\% & (${\pm}4$\%) & ${\pm}9$\% & (${\pm}14$\%)  \\
\hline
\end{tabular}
\end{table*}

Details of our results from analysis of the WiggleZ dataset (all uncertainties quoted at $68$\% confidence) are as follows:
\begin{itemize}
\item We present an MCMC method which removes window function convolution effects while coadding the individual observed power spectra of a galaxy survey observed over several disconnected regions.
\item Applying this method to WiggleZ data we find that the survey is able to probe modes at-and-beyond the scale of the turnover.
\item Using an MCMC approach, we have fit an asymmetric logarithmic parabola to the observed WiggleZ power spectrum.  From this analysis we find the scale of the turnover to be $k_{\rm 0}{=}0.0160^{+0.0035}_{-0.0041}$.  This is in excellent agreement with our fiducial standard \LCDM\ value of $k_{\rm 0}=0.016$ [Mpc/$h$].
\item Parameterising the power spectrum beyond the scale of the turnover as $P(k{<}k_{\rm 0}){\propto}k^n$, we find $n{>}{-}1$ at nearly $95$\% confidence.  The standard \LCDM\ value of $n{=}1$ is easily accommodated by our fit.
\item The continuance at large scales of the small-scale asymptotic value of $n{=}{-}3$ is completely ruled-out by our analysis.
\item We have performed the first measurement of the peak position in the cosmological power spectrum, representing the first secure and quantified observation of this feature to date.
\end{itemize}

From this measurement we then extract -- for the first time using a measurement of the turnover scale -- the following information:
\begin{itemize}
\item A model-independent distance measurement to $z{=}0.62$ in units of the $z{=}3145$ horizon scale at the redshift of matter-radiation equality ($r_{\rm H}$) of $D_{\rm V}(z_{\rm eff}{=}0.62){/}r_{\rm H}{=}18.3^{+6.3}_{-3.3}$.
\item A measurement of the cosmological density parameter $\Omega_{\rm M}h^2{=}0.136^{+0.026}_{-0.052}$ (assuming a $N_{\rm eff}{=}3$ prior).
\item A measurement of the redshift of matter-radiation equality $z_{\rm eq}{=}3274^{+631}_{-1260}$ (assuming a $N_{\rm eff}{=}3$ prior).
\end{itemize}

Looking to the future, we have computed forecasts for the turnover precision attainable by BOSS and Euclid.  We find that BOSS should substantially improve upon the results presented here, reaching precisions in $(k_{\rm 0},\Omega_{\rm M} h^2, z_{\rm eq},N_{\rm eff})$ of $({\pm}9,{\pm}10,{\pm}10,{}^{+78}_{-56})$\%.  This represents sufficient precision to sharpen the constraints on $N_{\rm eff}$ from WMAP, particularly in its upper limit.  For Euclid, we find corresponding attainable precisions of $({\pm}3,{\pm}4,{\pm}4,{}^{+17}_{-21})$\%.  This represents a precision approaching our forecasts for Planck.

We emphasise that these results are all obtained within the theoretical framework of Gaussian primordial fluctuations.  In the event that cosmological structure formation was seeded by non-Gaussian fluctuations, scale-dependent bias effects may substantially change the standard \LCDM\ predictions for $P(k)$ on scales comparable to and larger than the turnover \citep{Dalal:2008p1750,Desjacques:2009p1746,Desjacques:2011p1747,Desjacques:2011p1748,Giannantonio:2010p1749}.  Measurements of primordial Gaussianity from studies of large-scale structure have proven to provide constraints of comparable precision to those from the CMB \citep{Slosar:2008p1751} with forecasts for future surveys suggesting that the two approaches will remain complementary for some time \citep{Giannantonio:2012p1745}.  Interestingly, current constraints  allow for cases where the power spectrum exhibits no turnover but merely an inflection at turnover scales, with $P(k)$ being a declining function of $k$ at large scales.  Many such models are compatible with the results presented in this study but the lower limit we place on $n$ in the large-scale limit of $P(k){\propto}k^n$ suggests that the WiggleZ power spectrum measurement may be capable of constraining their allowed parameter ranges.  However, such effects are expected to scale with galaxy bias $b$ as $(b{-}1)$ \citep[\eg][]{Dalal:2008p1750} and WiggleZ galaxies are known to have a bias near unity \citep[\eg][]{Blake:2010p74}, greatly reducing the degree of such effects for this survey.  Regardless, we forgo this analysis for now and focus instead on the case of purely Gaussian primordial fluctuations.  Future studies involving WiggleZ bispectrum measurements will address this science.

Furthermore, while this paper focuses on the scale of the turnover as a cosmological constraint, there is much more information present in the full shape of the power spectrum which could provide (for example) constraints on $\Omega_{\rm M} h^2$ (see Parkinson \etal, submitted), tightening the constraints presented here not only on this parameter but on $N_{\rm eff}$ as well.  Such a situation is similar to that of BAO studies where final cosmological constraints tend to incorporate information from both the BAO scale and the shape of the underlying power spectrum \citep[see Figure 6 of][for example]{Blake:2011a}.  As such, the constraints presented here likely represent underestimates of the true potential of these experiments.

Lastly, with increased constraints on $N_{\rm eff}$ and $\Omega_{\rm M} h^2$, the absolute scale of the turnover could be accurately calibrated, permitting its use as a standard ruler for measuring the distance-redshift relation.  Our forecasts suggest that a regular volume of $V{\sim}10$ [(Gpc/h)$^3$] with a well defined effective redshift would permit a distance measurement of roughly $10$\% accuracy.  Euclid for instance could provide several such volumes arranged across redshift, perhaps enabling the first measurement of a Hubble diagram using the turnover scale as a standard ruler.  Given the purely linear evolution of the matter power spectrum on scales of the turnover, this could provide a powerful check against systematic redshift-dependent biases in BAO studies.

\section*{Acknowledgements}
We thank Francesco Montesano for his constructive and insightful examination of our manuscript.  We acknowledge financial support from the Australian Research Council through Discovery Project grants DP0772084 and DP1093738 and Linkage International travel grant LX0881951. GBP thanks Simon Mutch for his help with developing the MCMC code used for the analysis in this study.  CB acknowledges the support of the Australian Research Council through the award of a Future Fellowship.  SC and DC acknowledge the support of Australian Research Council QEII Fellowships. MJD and TMD thank the Gregg Thompson Dark Energy Travel Fund for financial support. GALEX (the Galaxy Evolution Explorer) is a NASA Small Explorer, launched in 2003 April. We gratefully acknowledge NASAÕs support for construction, operation and science analysis for the GALEX mission, developed in cooperation with the Centre National dÕEtudes Spatiales of France and the Korean Ministry of Science and Technology. We thank the Anglo-Australian Telescope Allocation Committee for supporting the WiggleZ survey over nine semesters, and we are very grateful for the dedicated work of the staff of the Australian Astronomical Observatory in the development and support of the AAOmega spectrograph, and the running of the AAT.  We are also grateful for support from the Centre for All-sky Astrophysics, an Australian Research Council Centre of Excellence funded by grant CE11000102.

\bibliographystyle{mn2e}
\bibliography{biblio}

\end{document}